\documentclass[11pt]{article}

\usepackage{amsmath,amssymb}
\usepackage{latexsym}
\usepackage{graphicx}
\usepackage{bm}

\newtheorem{thm}{Theorem}%[section]

\newtheorem{rem}{Remark}

\def\Image{\mathop{\rm Im}}
\def\Real{\mathop{\rm Re}}

\def\diam{\mathop{\rm diam}}

\begin{document}
\title{Preparing materials with a desired refraction coefficient}

\author{A. G. Ramm$\dag$\footnotemark[1]
\\
\\
$\dag$Mathematics Department, Kansas State University,\\
Manhattan, KS 66506-2602, USA
}

\renewcommand{\thefootnote}{\fnsymbol{footnote}}
\footnotetext[1]{Email: ramm@math.ksu.edu}

\date{}
\maketitle

\begin{abstract}
A recipe is given for creating material with a desired refraction 
coefficient by embedding many small particles in a given material.
To implement practically this recipe, some technological problems are 
to be solved. These problems are formulated.

\end{abstract}



\section{Introduction}
In a series of papers \cite{R508}-\cite{R523} and in the 
monograph \cite{R476}
the author has developed wave scattering theory by many small bodies of
arbitrary shapes and, on this basis, proposed a method for creating materials
with some desired properties. The goal of this paper is to explain this method
and to formulate two nanotechnological engineering problems that 
should be solved
in order that the proposed method can be implemented practically.

Let us formulate the statement of the problem. 
Let $D\subset \mathbb{R}^3$ be a bounded domain
filled in by a material with a known refraction coefficient. The scalar wave scattering
problem consists of finding the solution to the Helmholtz equation:
\begin{equation}
\label{eq1}
L_0u:=(\nabla^2+k^2n_0^2(x))u=0 \quad \text{in}\quad \mathbb{R}^3,\quad\Image 
n_0^2(x)\ge 0,
\end{equation}

\begin{equation}
\label{eq2}
u=u_0+v, \quad u_0:=e^{ik\alpha\cdot x},
\end{equation}

\begin{equation}
\label{eq3}
v=A_0(\beta,\alpha)\frac{e^{ikr}}{r}+ o(\frac{1}{r}),
\quad r:=|x|\to\infty,\quad \beta:=\frac{x}{r}.
\end{equation}
The function $n_0(x)$ is assumed bounded, piecewise - continuous,
\begin{equation}
\label{eq4}
n(x)=1 \quad \text{in}\quad D':=\mathbb{R}^3\backslash D,\quad \Image n_0^2(x)\ge 0.
\end{equation}
The function $A(\beta,\alpha)$ is called the scattering amplitude. 
The wavenumber $k=\frac{2\pi}{\lambda}$, where $\lambda$ is the wavelength in $D'$,
$\alpha\in S^2$ is the direction of the incident plane wave $u_0$, $S^2$ is the unit 
sphere in $\mathbb{R}^3$, $\beta\in S^2$ is the direction of the scattered wave. 
The solution to problem \eqref{eq1}-\eqref{eq3} is called the scattering solution. 
It is well-known (\cite{R190}) that this solution exists and is unique under
our assumptions, namely $n_0^2(x)$ is bounded 
$\sup_{x\in\mathbb{R}^3}|n_0^2(x)|\le n_0=const$, $\Image n_0^2(x)\ge 0$, 
$n_0^2(x)=1$ in $D'$, so that $A_0(\beta,\alpha)$ is determined uniquely if $n_0^2(x)$
is given. We assume $k>0$ fixed and do not show the dependence of $A_0(\beta,\alpha)$
on $k$. The operator $L_0$ at a fixed $k>0$ can be considered as a Schroedinger 
operator
$L_0=\nabla^2+k^2-q_0(x)$, where $q_0(x):=k^2n_0^2(x)-k^2$.

{\bf\underline{Problem(P)}}:{\it We want to construct a material in $D$ with a desired
refraction coefficient, i.e., with a desired function $n^2(x)$, $\Image n^2(x)\ge 0$.}

Why is this problem of practical interest? We give just two reasons 
(out of, possibly, many). 
First, creating a refraction coefficient such that the corresponding 
material has {\it negative refraction} is of practical interest. One says that
a material has negative refraction if the group 
velocity in this material is directed 
opposite to the phase velocity. 

Secondly,  creating such a refraction coefficient that the corresponding
scattering amplitude $A(\beta)=A(\beta,\alpha)$ at a fixed $\alpha\in S^2$ 
(and a fixed $k>0$)
approximates with any desired accuracy $\epsilon>0$ an arbitrary given 
function $f(\beta)\in L^2(S^2)$ is of practical interest. 
This problem we call the problem of creating material with a desired 
wave-focusing property. 

In Section 2 we explain how to create such a material. In Section 3 we discuss the technological 
aspects of our recipe for creating material with a desired refraction coefficient. Creating material
with negative refraction is discussed in Section 4. 

\section{Creating material with a desired refraction coefficient}

The basic idea of our method is to embed in $D$ many small particles 
$D_m$, $1\leq m \leq M$, with a prescribed boundary impedance. 
The number of the embedded particles in an arbitrary open subdomain 
$\Delta \subset D$ is given
asymptotically as 

\begin{equation}
\label{eq5}
\mathcal{N}(\Delta)=\frac{1}{\phi(a)}\int_\Delta N(x)dx [1+o(1)],\qquad 
a\to 0.
\end{equation}
Here $\phi >0$ if $a>0$, $\phi(0)=0$, $\phi$ is strictly monotonically growing,
$N(x)\ge 0$ is a piecewise-continuous function which we can choose as we want. 
The smallness of the embedded particles means that $ka\ll 1$, where 
$a=\frac{1}{2}\max_m\diam D_m$, where $D_m$ is the $m$-th small particle. 
Let $M=M(a)$ be the total number of the embedded particles. For simplicity
we assume that $D_m$ is a ball, centered at a point $x_m\in D$, of radius $a$.
The distance $d:=\min_{m\not= j}dist(D_m,D_j)$ between neighboring particles is much 
larger than $a$, $d\gg a$, 
but $d$ {\it is not assumed much larger than $\lambda$}, so in our theory
there can be many small particles on the wavelength in $D$. This means
that the distribution of the embedded particles {\it is not necessarily "diluted".}

The scattering problem in the region with embedded particles is formulated 
as follows:
\begin{equation}
\label{eq6}
L_0 U=0 \quad \text{in}\quad \mathbb{R}^3\backslash \bigcup_{m=1}^M D_m,
\end{equation}

\begin{equation}
\label{eq7}
U_N=\zeta_mU \quad \text{on}\quad S_m, \quad 1\le m\le M,
\end{equation}

\begin{equation}
\label{eq8}
U(x)=u(x)+V,\quad \frac{\partial V}{\partial r}-ikV=o(\frac{1}{r}), \quad r\to\infty.
\end{equation}
Here $N$ is the unit normal to $S_m$, pointing out of $D_m$, 
$S_m$ is the surface of $D_m$, $M$ is the total number of the 
embedded particles $D_m$, $\zeta_m$ 
is the boundary impedance, $\Image \zeta_m\le 0$, 
$u$ is the scattering solution in the absence of the embedded particles, i.e.,
the solution to problem \eqref{eq1}-\eqref{eq3}.

Let us assume that
\begin{equation}
\label{eq9}
 \zeta_m=\frac{h(x_m)}{a^\varkappa}, \quad -1<\varkappa <2,\quad \varkappa 
= const,
\end{equation}
where $h(x)$ is a piecewise-continuous function in $D$,

\begin{equation}
\label{eq10}
\Image h(x) \le 0.
\end{equation}
The function $N(x)\ge 0$ in \eqref{eq4} and $h(x)$, $\Image h\le 0$,
we can choose {\it as we wish}. These functions will define the desired 
refraction 
coefficient of the new material which is obtained in the limit $a\to 0$.
We assume that 

\begin{equation}
\label{eq11}
M=M(a)=O(a^{\varkappa-2}),\quad d=O(a^{\frac{2-\varkappa}{3}}),
\end{equation}
and the function $\phi(a)$ in \eqref{eq4} is

\begin{equation}
\label{eq12}
\phi(a)=a^{2-\varkappa}.
\end{equation}

The solution to problem \eqref{eq5}-\eqref{eq7} depends on $a$, $U=U(x,a)$,
but we do not show this dependence for brevity. We also do not show 
the dependence of $U$ on $\alpha$ and $k$, because they are fixed.
Our standing assumptions are \eqref{eq4}, \eqref{eq5}, \eqref{eq9}--\eqref{eq12}.
They are not repeated in the statements of Theorems. 

Denote by $g(x,y)$ the Green's function of the operator $L_0$
in the absence of the embedded particles, $L_0g(x,y)=-\delta(x-y)$ in $R^3$,
$g$ satisfies the radiation condition.

\begin{thm}
\label{theorem1}
Problem \eqref{eq5}-\eqref{eq7} has a solution, this solution is unique, and it is of the form:
\begin{equation}
\label{eq13}
U(x)=u(x)+\sum_{m=1}^M\int_{s_m} g(x,s)\sigma_m(s)ds,
\end{equation}
where the function $\sigma_m$ are uniquely determined by the data, i.e., by 
$u_0(x,\alpha)$, $\zeta_m$, $D_m$, $1\le m\le M$, and $n_0^2(x)$.
\end{thm}

\begin{thm}
\label{theorem2}
The solution $U$ to problem \eqref{eq6}--\eqref{eq8} has the following asymptotic as $a\to 0$
\begin{equation}
\label{eq14}
U(x) = u(x) + \sum_{m=1}^{M(a)} g(x,x_m)Q_m + o(1),\quad a\to 0,\quad
\inf_m |x-x_m|\ge d,
\end{equation}
where
\begin{equation}
\label{eq15}
Q_m:=\int_{S_m}\sigma_m(s)ds=-4\pi h(x_m)u_e(x_m)a^{2-\varkappa}[1+o(1)],\quad a\to 0,
\end{equation}
\begin{equation}
\label{eq16}
\sigma_m (s) = -h(x_m)u_e(x_m) a^{-\varkappa}[1+o(1)],\quad a\to 0,
\end{equation}
and
\begin{equation}
\label{eq17}
u_e(x_m):=u_e^{(m)}(x_m),\quad u_e^{(m)}(x):=u(x)+ \sum_{m'\not= m}\int_{S_{m'}}g(x,s)
\sigma_{m'}(s)ds.
\end{equation}
\end{thm}

\begin{thm}
\label{theorem3}
Assume that $f$ is an arbitrary bounded piecewise continuous function in $D$, and
the points $x_m$ are distributed in $D$ so that \eqref{eq5} holds. Then
\begin{equation}
\label{eq18}
\lim_{a\to0} \phi(a)\sum_{m=1}^{M(a)}f(x_m)= \int_D f(x)N(x)dx.
\end{equation}
\end{thm}

\begin{thm}
\label{theorem4}
Under the assumptions \eqref{eq4}, \eqref{eq5}, \eqref{eq9}--\eqref{eq12}
the following limit exists:
\begin{equation}
\label{19}
\lim_{a\to0} u_e^{(m)}(x) = u_e(x),\quad \forall x\in D.
\end{equation}
This limit does not depend on $m$ and is the unique solution to the equation
\begin{equation}
\label{eq20}
u_e(x) = u(x) + \int_D g(x,y)p(y) u_e(y)dy,
\end{equation}
where
\begin{equation}
\label{eq21}
p(x) = 4\pi N(x)h(x).
\end{equation}

\end{thm}

\begin{thm}
\label{theorem5}
The limiting material in $D$, which is obtained as $a\to 0$,
is described by the equation 
\begin{equation}
\label{eq22}
Lu_e = 0\quad \text{in}\quad D,\quad L:=\nabla^2+k^2n^2(x):=\nabla^2+k^2-q(x),
\end{equation}
\begin{equation}
\label{eq23}
n^2(x):=n_0^2(x) - k^{-2}p(x),\quad p(x):=4\pi N(x)h(x),\quad q(x):=q_0(x)+p(x),
\end{equation}
\begin{equation}
\label{eq24}
u_e=u(x)+v_e(x),
\end{equation}
where $v_e(x):=\int_Dg(x,y)p(y)u_e(y)dy$.

The scattering amplitude, corresponding to $n^2(x)$, 
(or, which is equivalent to the potential $q(x)$) is:
\begin{equation}
\label{eq25}
A(\beta,\alpha) = A_0(\beta,\alpha)+\frac{1}{4\pi}\int_D u(y,-\beta)p(y)u_e(y)dy,
\end{equation}
where $u(x,-\beta)$ is the solution to problem \eqref{eq1}--\eqref{eq3} with
$u_0=e^{-ik\beta \cdot x}$, that is, the scattering solution in the absence of the
embedded particles when the incident direction of the plane wave $u_0$ equals to $-\beta$.
\end{thm}

\begin{rem}{\rm
\label{remark1}
Formula \eqref{eq25} is based on the Ramm's lemma (see \cite[p. 46]{R190}) which says
\begin{equation}
\label{eq26}
g(x,y) = \frac{e^{ik|y|}}{4\pi |y|}u(x,-\beta) + O(\frac{1}{|y|^2}),\quad
|y|\to\infty,\quad \beta:=\frac{y}{|y|},
\end{equation}
where $u(x,-\beta)$ is the scattering solution with the incident direction $-\beta\in S^2$
and the remainder $ O(\frac{1}{|y|^2})$ is uniform with respect to $x$ running 
through any fixed bounded domain.
}
\end{rem}

We can now formulate {\it a recipe for creating material with a desired refraction coefficient:}

{\bf \underline{Recipe}: To obtain material with a desired refraction coefficient $n^2(x)$,
given a bounded domain $D$, filled with a material with the coefficient $n_0^2(x)$,
one embeds into $D$ small balls $D_m$, centered at $x_m$ and of radius 
$a$, distributed 
according
to \eqref{eq5}, with $\phi(a)$ defined in \eqref{eq12}, so that conditions
 \eqref{eq9}-\eqref{eq11} hold. Then one  finds $h(x)$ and $N(x)$ from the equation 
\begin{equation}
\label{eq27}
k^2[n_0^2(x) - n^2(x)]:=p(x)=4\pi h(x)N(x).
\end{equation}
Equation \eqref{eq27} for $N(x)\ge 0$ and $h(x), \Image h\le 0$, where
$p(x)$ is given, has infinitely many solutions. For example, one can choose 
$N(x)>0$ arbitrary, and then find uniquely $h_1$ and $h_2$ by the formulas:
\begin{equation}
\label{eq28}
h_1:= \Real h = \frac{p_1}{4\pi N(x)},\quad h_2:= \Image h = \frac{p_2}{4\pi N(x)}
\end{equation}
where $p = p_1+ i p_2$,\, $p_1=\Real p$,\, $p_2=\Image p$.

By Theorem~\ref{theorem5} the material with the embedded small particles has 
the desired coefficient $n^2(x)$ with the error that goes to zero as $a\to 
0$.
}

We do not give proofs of Theorem~\ref{theorem1}--\ref{theorem5}. These proofs can be 
found in the cited papers. Our goal is to formulate clearly the recipe for 
creating materials with a desired $n^2(x)$, so that engineers and physicists 
can try to implement this recipe practically.  

\section{A discussion of the recipe}

There are two technological problems that should be solved in order that the recipe 
be implemented practically.

{\bf Problem 1:} {\it How does one embed many small particles in a given material 
so that the desired distribution law \eqref{eq5},\eqref{eq11}-\eqref{eq12} 
is satisfied?}

{\bf Problem 2:} {\it How does one prepare a small particle with the desired boundary 
impedance $\zeta_m = \frac{h(x_m)}{a^{\varkappa}}$?}

The first problem, possibly, can be solved by stereolitography. The second problem one
should be able to solve because the limiting cases $\zeta_m=0$ (hard particles) and 
$\zeta_m=\infty$
(soft particles) can be prepared, so that any intermediate value of $\zeta_m$ one 
should be able to
prepare as well. The author formulates the above technological problems in the hope that
engineers get interested and solve them practically. 

\subsection{Negative refraction}

Material with negative refraction is, by definition, a material in which 
group velocity 
is directed opposite to the phase velocity, (see \cite{A} and references therein).
Group velocity is defined by the formula $\mathbf{v}_g = 
\nabla_\mathbf{k}\omega(\mathbf{k})$.  Phase 
velocity 
$\mathbf{v}_p$ is directed along the wave vector 
$\mathbf{k}^0=\frac{\mathbf{k}}{|\mathbf{k}|}$.
In an isotropic material $\omega = \omega(\mathbf{|k|})$, and $\omega = 
\frac{c|\mathbf{k}|}{n(x,\omega)}$,
so $\omega n(x,\omega )=c|\mathbf{k}|$. Differentiating this equation yields
\begin{equation}
\label{eq29}
\nabla_{\mathbf{k}}\omega \big{[}n(x,\omega)+ \omega \frac{\partial n}{\partial 
\omega }\big{]} 
= c\mathbf{k}^0.
\end{equation}
Thus,
\begin{equation}
\label{eq30}
\mathbf{v}_g \big{[}n+\omega \frac{\partial n}{\partial \omega }\big{]} = 
c\mathbf{k}^0.
\end{equation}
Wave speed in the material with refraction coefficient $n(x,\omega )$ is 
$|\mathbf{v}_p|=\frac{c}{n(x,\omega )}$, where $c$ is the wave speed in vacuum (in 
$D'$).

For $\mathbf{v}_g$ to be directed opposite to $\mathbf{k}^0$, that is, 
opposite to $\mathbf{v}_p$, it is necessary and sufficient that 
\begin{equation}
\label{eq31}
n+\omega \frac{\partial n}{\partial \omega } < 0.
\end{equation}
If the new material has $n(x,\omega)$ satisfying \eqref{eq31}, then the new 
material has negative refraction.

One can create material with $n(x,\omega )$ satisfying \eqref{eq31} by choosing 
$h=h(x,\omega )$ properly.
Namely, \eqref{eq27} implies
\begin{equation}
\label{eq32}
n^2(x,\omega ) = n_0^2(x) - 4\pi k^{-2} N(x) h(x,\omega).
\end{equation}
Assuming $n_0^2> 4\pi k^{-2} N(x) h(x,\omega)$,\, $\Image h = 0$,\, one has
\eqref{eq31} satisfied if 
$$
\sqrt{n_0^2 - 4\pi k^{-2}N(x)h(x,\omega )} < \frac{2\pi k^{-2}N(x)\frac{\partial 
h(x,\omega )}{\partial \omega }}{\sqrt{n_0^2 - 4\pi k^{-2}N(x)h(x,\omega )}},
$$
or
\begin{equation}
\label{eq33}
t(x):=\frac{n_0^2(x)}{2\pi k^{-2}N(x)} < \frac{\partial h}{\partial \omega } + 2h.
\end{equation}
This inequality has many solutions, e.g., the function $h(x,\omega ) = 
e^{-2\omega }h_0 + 
\frac{1}{2}e^{2\omega }t(x)$ solves \eqref{eq33}.

Let us formulate the technological problem solving of which allows one to implement practically our method for creating materials with negative refraction.

{\bf Problem 3:} {\it How does one prepare a small particle with the impedance
$\zeta_m=\frac{h(x_m,\omega )}{a^{\varkappa}}$, where $h(x,\omega )$ satisfies 
\eqref{eq33}?}

\end{document}